\newskip\oneline \oneline=1em plus.3em minus.3em
\newskip\halfline \halfline=.5em plus .15em minus.15em
\newbox\sect
\newcount\eq
\newbox\lett

\def\simlt{\mathrel{\lower2.5pt\vbox{\lineskip=0pt\baselineskip=0pt
           \hbox{$<$}\hbox{$\sim$}}}}
\def\simgt{\mathrel{\lower2.5pt\vbox{\lineskip=0pt\baselineskip=0pt
           \hbox{$>$}\hbox{$\sim$}}}}

\newdimen\short
\def\adv{\global\advance\eq by1}
\def\set#1#2{\setbox#1=\hbox{#2}}
\def\nextlet#1{\global\advance\eq by-1\setbox
                \lett=\hbox{\rlap#1\phantom{a}}}

\newcount\eqncount
\eqncount=0
\def\equn{\global\advance\eqncount by1\eqno{(\the\eqncount)} }
\def\put#1{\global\edef#1{(\the\eqncount)}           }

\def\np{{\it Nucl. Phys.}}
\def\pl{{\it Phys. Lett.}}

\def\cWW{1}
\def\cAth{2}
\def\cSch{3}
\def\cAnt{4}
\def\cCamb{5}
\def\cConf{6}
\def\cGC{7}
\def\celn{8}
\def\cBarb{9}
\def\ck{10}
\def\cil{11}
\def\caekn{12}
\def\clep{13}
\def\cmsu{14}
\def\cas{15}
\def\cbl{16}

\magnification=1200
\hsize=6.0 truein
\vsize=8.5 truein
\baselineskip 14pt

\nopagenumbers

\rightline{CPTH-A193.0892}
\vskip 1.0truecm
\centerline{\bf FRACTIONALLY CHARGED PARTICLES}
\centerline{\bf AND}
\centerline{\bf SUPERSYMMETRY BREAKING IN 4D STRINGS}
\vskip 1.0truecm
\centerline{{\bf I. Antoniadis} and {\bf K. Benakli}}
\vskip .5truecm
\centerline{\it Centre de Physique Th{\'e}orique}
\centerline{\it Ecole Polytechnique, 91128 Palaiseau, France}
\vskip 2.5truecm
\centerline{\bf ABSTRACT}
\vskip .5truecm
Four-dimensional string theories predict in general the existence of
light exotic particles with fractional electric charges. Such particles
could escape present observations if they are confined by a gauge group of
the ``hidden" sector into integrally charged states. It is conceivable
that the same gauge group is also responsible for dynamical supersymmetry
breaking, via gaugino and scalar condensation. The communication of the
breaking to the observable sector is now mediated by ordinary gauge
interactions, implying that the confining scale can be in the TeV-region.

We study the main phenomenological implications of this possibility. In
particular, we analyze the pattern of supersymmetry breaking and the
mass-spectrum of the sparticles. We also show that this scenario can be
consistent with the unification of all coupling constants at the string
scale.
\vskip 3.5truecm
\noindent CPTH-A193.0892       \hfill\break
\noindent August 1992
\vfill\eject

\footline={\hss\tenrm\folio\hss}\pageno=1

A main issue in string theory is to derive a four-dimensional model,
which leads, at low energies, to the Standard Model of electroweak
and strong interactions.
Although there is a lot of arbitrariness due to the infinite vacuum
degeneracy, there are still many stringy constraints emerging from the
internal consistency of the theory. Here, we discuss two main implications
related to the generic existence of particles with fractional electric
charges [\cWW-\cSch] and to the problem of dynamical supersymmetry
breaking.

One important restriction is on the allowed matter representations.
Given a simple gauge group factor $G_i$, the dimensions of the possible
unitary massless representations are constrained by the level $k_i$ of
the associated Kac-Moody algebra on the world-sheet. The positive integer
paremeter $k_i$  determines also the corresponding tree-level gauge
coupling constant  $\alpha_i$, in terms of the four-dimensional string
coupling $\alpha$, $\alpha_i = \alpha / k_i$. Models with $k_i>1$ are
difficult to  construct and they lead in general to ``exotic"
representations
in the  massless spectrum, e.g. $SU(3)$-octets or $SU(2)$-triplets or
higher. In contrast, models with $k_i=1$ are simpler, and they guarantee
that the only possible massless  representations are singlets or doublets of
$SU(2)$  and singlets or triplets of $SU(3)$. More generally, they allow
vectors and spinors  of orthogonal groups or antisymmetric representations
of unitary  groups. Moreover, they imply an automatic unification of all
couplings at the Planck scale $M_p$. However, in this case, it is
impossible  to impose the value for the weak angle $sin^2{\theta}_w =
{3\over 8}$ at $M_p$ similtaneously with the observed electric charge
quantization at the level of the string-derived Standard Model [\cSch].
Thus, unless the Weinberg angle at $M_p$ is too small, we must deal with
fractional electric charged particles (FEC-particles). In general, a weaker
charge quantization condition can be imposed depending on the
compactification; for instance, that color neutral states have charges
multiples of  $1/2$ for fermionic constructions, or $1/N$ for $Z_N$
orbifolds [\cAnt].

The lightest FEC-particle must be stable. Previous searches of such
particles lead to several bounds on their masses provided their charges are
not ridiculously small [\cCamb]. On the one hand, masses below
approximately $100{\rm GeV}$ are excluded by accelerator experiments. On
the other hand, masses above a few TeV would overclose the Universe and are
also excluded, leaving a  ``window" for masses of a few hundred GeV.
However, for these masses, an estimation of their relic abundancies seems
to
contradict the upper  bounds found in Millikan experiments by several
orders
of magnitude  [\cAth]. Therefore, free light FEC-particles are excluded,
unless there is some unkown  mechanism which supresses their abundance on
earth. There remain two possible ways to escape the experimental
constraints. The first is to make them superheavy, with masses of the order
$M_p$. The second and most natural is to confine them by a gauge group of
the hidden sector into integrally charged bound states like the quarks in
QCD [\cConf].

Another important problem is that of supersymmetry breaking, which must
occur at energies of the order of the weak-interaction scale, to protect
the gauge hierarchy. A popular scenario for its attractive  features is
based on gaugino condensation in a gauge group of the  hidden sector, at a
high scale $\Lambda \sim 10^{13}{\rm GeV}$ [\cGC]. The supersymmetry
breaking is communicated to the observable sector by gravitational
interactions, giving rise to small mass-splittings  for the superpartners
of
the order of ${{\Lambda}^3 \over{M_p^2}} \sim 1{\rm Tev}$.

In this letter, we explore the idea that the gauge group $G$ which confines
the fractionally charged particles is also responsible for the breaking of
supersymmetry. In any case, the existence of a group which confines the
FEC-particles could always provide an additional source of supersymmetry
breaking via non-perturbative condensate formation of gauginos or
scalar components of chiral supermultiplets. The latter could give rise to
mass-splittings in the ``hidden" sector, in particular between the
components of the fractionally charged matter supermultiplets, of the order
of the confinement scale $\Lambda$ through dimension-four renormalizable
interactions. Since the breaking is now communicated to the ``observable"
sector by strong and electroweak interactions, $\Lambda $ must be of the
order of a few TeV. This may be a generic feature of a large class of
models. In this work, we study the main phenomenological implications of
this mechanism. In particular, we analyze the novel pattern of
supersymmetry
breaking, the mass-spectrum of the sparticles and the modifications to the
coupling constant unification.  This scenario also predicts a rich spectrum
of new composite particles in the TeV-region, called ``cryptons" [\celn].
\vskip 1.0cm
\centerline{\bf SPARTICLE MASSES}
\vskip 0.5cm

To simplify the discussion, we restrict our analysis to the following
content
of the ``hidden" sector:
\par\noindent
-The gauge group $G$ is $SO(N)$ or $SU(N)$.
\par\noindent
-The FEC-particles are in vector representations of $G$: $N$ and
$\overline{N}$ of $SU(N)$ or $N$ of $SO(N)$ with a common mass-splitting
$\delta M$ between scalars and fermions superpartners.
Moreover, these particles transform non trivially under the Standard Model
gauge group. The allowed representations are denoted by $Q\equiv (3,2,{e_Q
\over 3N})$, $D\equiv (\bar{3},1, {e_D \over 3N})$, $L\equiv (1,2,{e_L
\over
N})$, $E\equiv (1,1, {e_E \over N})$, under $SU(3)\times SU(2)$ with the
third position inside the  brackets denoting the electric charge (in the
case of doublet it is the charge of the upper component). The
numbers $e_{Q,D,L,E}$ must be integers, so that gauge singlet bound states
of
$G\times SU(3)$ are  integrally charged. We also allow for different
multiplicities $n_Q$, $n_D$, $n_L$ and $n_E$ associated to the pairs of
chiral-antichiral representations $Q+{\bar Q}$, $D+{\bar D}$, $L+{\bar L}$
and $E+{\bar E}$, respectively.

Supersymmetry breaking is communicated to the observable sector through
gauge interactions. In fact, radiative corrections in the sparticle
propagators involving the exhange of FEC-fields induce mass-splittings
in the usual Standard Model supermultiplets. Below, we evaluate these
contributrions to the lowest order.

We first consider the gaugino masses $m_{\tilde g_i}$, which receive
contributions from the one-loop graphs involving the
exchange of the fermion and scalar components of all FEC-supermultiplets.
They were computed in Ref.[\cBarb] with the result:
$$
m_{\tilde{g}_i}= {\sum}_j c_{\tilde g_i} M_j F({\delta M \over M_j}),
\equn\put\mgeg
$$
where $c_{\tilde g_i}$ is a constant factor depending on the gauge group
and
matter representations, while $F$ is a function which depends only on the
ratio of the mass-splitting $\delta M$ over the supersymmetric mass $M_j$
of
the $j$-th FEC-multiplet:
$$
F(x)= {-{{(1+x)^2} \over {1-(1+x)^2}} \ln ({(1+x)^2})}+ {{{(1-x)^2}
\over {1-(1-x)^2}} \ln ({(1-x)^2})}.
\equn\put\intf
$$
The function ${1 \over x} F({x})$ is nearly constant for $x\simlt 1$ and it
goes to zero asymptotically as $1/ x$ for $x\simgt 2$. In order to
avoid too small gaugino masses, we concentrate on the region $M_j \simgt
\delta M$, where we can use the following approximation :
$$
M_j F({{\delta M} \over {M_j}}) \simeq 2  \delta M.
\equn\put\mapro
$$
The gaugino masses are then independent of the supersymmetric masses of the

FEC-particles and they are proportional to the mass-splitting ${\delta M}$.

The explicit formulas for gluinos, wino and the two neutralinogauginos soft
 masses are:

$$
\eqalign{ m_{\tilde g_3}&= {\alpha_s \over 8 \pi}
4~N~\delta M ~(2 n_Q + n_D) \cr
 m_{\tilde g_2} &={\alpha_{em} \over 4 \pi} 2~N {\delta M \over
{\sin^2{\theta}_W}} (3 n_Q + n_L ) \cr \cr
m_{\tilde g_1} &={\alpha_{em} \over 4 \pi}{4\over N}{\delta M \over
{\cos^2{\theta}_W}}(n_Q {{(2e_Q-3N)^2} \over 6}+
n_D {{e_D^2} \over 3} +n_L{{(2e_L -N)^2} \over 2 }+ n_E e_E^2 ). \cr}
\equn\put\mgau
$$

We now consider squark and slepton masses. These are in general model
dependent because they also depend on possible superpotential terms among
ordinary matter and FEC-particles. To get a rough estimate for the various
orders of magnitude, we calculate the contributions
to scalar masses from radiative corrections of gauge interactions.
The contribution of tadpoles, which involve the exchange of FEC-scalars
coming from the $U(1)$-hypercharge D-term, vanishes in our case, since we
consider left-right symmetric representations with a common mass-splitting
$\delta M$.The leading contribution are two-loop diagrams involving one loop of Standard 
Model particles and one loop of FEC
particles [\ck]. 
These diagrams are logarithmically divergent and require the use of some
ultraviolet cut-off, corresponding to the higher scale up to which
the effective field theory remains valid. Replacing the cut-off with the
string unification scale $M_{\rm SU}$, the computation of radiative
corrections to the squark and slepton masses give [\cil]:
$$
\eqalign {\delta m_{\tilde u_L}^2&={4 \over 3} m_3^2
+{3 \over 4}  m_2^2 + {1 \over 36} m_1^2   \cr
\delta m_{\tilde d_L}^2&={4 \over 3} m_3^2 +{3 \over 4} m_2^2 +
{1 \over 36} m_1^2   \cr
\delta m_{\tilde u_R}^2&={4 \over 3} m_3^2 + {4 \over 9} m_1^2 \cr
\delta m_{\tilde d_R}^2&={4 \over 3} m_3^2 + {1 \over 9} m_1^2 \cr
\delta m_{\tilde e_L}^2&={3 \over 4} m_2^2 + {1 \over 4} m_1^2 \cr
\delta m_{\tilde \nu _L}^2&={3 \over 4} m_2^2 + {1 \over 4} m_1^2 \cr
\delta m_{\tilde e_R}^2&=m_1^2 ,}
\equn\put\mscal
$$
with
$$
\eqalign{ 
m_1^2 &=({\alpha_{em} \over  \pi})^2 {1 \over  N^3} {{\delta M^2} \over \cos^4 
{\theta_W}} (
n_Q {(2 e_Q-3N)^4 \over 216}\ln ({M_{\rm SU} \over M_Q}) +
n_D {e_D^4 \over 27}\ln ({M_{\rm SU} \over M_D}) \cr & +  
n_L {(2 e_L-N)^4 \over 8}\ln ({M_{\rm SU} \over M_L}) +
 n_E e_E^4 \ln ({M_{\rm SU} \over M_E}))\cr 
m_2^2 &=({\alpha_{em} \over  \pi})^2 N {{\delta M^2} \over \sin^4 {\theta_W}} (
3 n_Q \ln ({M_{\rm SU} \over M_Q}) +  
n_L \ln ({M_{\rm SU} \over M_L})) \cr
m_3^2 &=({\alpha_s \over  \pi})^2 N \delta M^2  (
2 n_Q \ln ({M_{\rm SU} \over M_Q}) +  
n_D \ln ({M_{\rm SU} \over M_D})) \cr}
\equn\put\md
$$
where $\alpha_i$,
$i=1,2,3$ are the Standard Model coupling constants of $U(1)$, $SU(2)$ and
$SU(3)$, respectively. In {\mscal} we have not included the contribution
coming from the Higgs sector, which is model dependent and will be
discussed below. Moreover, in the absence of Yukawa couplings, the scalar
masses are the same for all families. Numerical examples for the above
sparticle masses will be given at the end.
\vskip 1.0cm
\centerline{\bf COUPLING CONSTANT UNIFICATION}
\vskip 0.5cm

The existence of light fractionally charged particles could alter
significantly the coupling constant unification. In fact, at the one
loop level, the gauge couplings at the energy scale $Q$ depend on the
particle content through the renormalization group formula:
$$
{1 \over {\alpha_i (Q)}} = {1 \over {\alpha_{\rm GUT}}}
+{b_i \over {2\pi}} {\ln ({M_{\rm GUT} \over {Q}})},
\equn\put\fbetag
$$
where $M_{\rm GUT}$ is the unification scale, and $b_i$ is the one-loop
beta-function
$$
b_i= -3 C(G_i) + \sum_{{\rm reps} R_i} T(R_i).
\equn\put\fbi
$$
In {\fbi} $C(G_i)$ is the quadratic Casimir of the group $G_i$ and equals
$N$ for $SU(N)$ and $N-2$ for $SO(N)$. $T(R_i)$ is the index of the
corresponding matter representation $R_i$; it is equal to ${1 \over 2}$ for
chiral supermultiplets in the fundamental representation of $SU(N)$, while
it is reduced to the sum of the squares of charges in the case of
$U(1)$-hypercharge. In the case of $SU(3)\times SU(2)\times U(1)$, one can
solve for $M_{\rm GUT}$ and $\sin^2{\theta}_W(m_{Z})$ as a function of the
strong and electromagnetic coupling constants $\alpha_s (m_{Z})$ and
$\alpha_{em} (m_{Z})$ at $m_Z$:
$$
\eqalign{\ln ({{M_{\rm GUT}} \over {M_Z}}) &={\pi \over 6}
({3 \over 5 \alpha_{em} (m_{Z})}- {8 \over 5 \alpha_s (m_{Z})})-0.21
- {1 \over 20}[{L_Q \over 6}(({{2e_Q -3N}\over N})^2 -7) \cr
&\qquad \qquad \qquad \qquad \qquad \qquad \qquad \qquad \qquad
+ {L_D \over 3}({e_D^2 \over N^2}-4) \cr
&\qquad \qquad \qquad \qquad \qquad \qquad \qquad \qquad \qquad
+{L_L \over 2}({{(2e_L -N)^2} \over N^2}+1) \cr
&\qquad \qquad \qquad \qquad \qquad \qquad \qquad \qquad \qquad
+ L_E ({e_E^2 \over N^2})],}
\equn\put\fmgut
$$
$$
\eqalign{
\sin^2{\theta}_W(m_Z ) &= 0.0029 + {1 \over 5}+{{7\alpha_{em} (m_{Z})}
\over {15 \alpha_s (m_{Z})}}+{{\alpha_{em} (m_{Z})}
\over {20 \pi}} [{L_Q \over 3}(-({{2e_Q -3N}\over N})^2 +22) \cr
&\qquad \qquad \qquad \qquad \qquad \qquad \qquad \qquad \qquad
- L_D (2({e_D^2 \over 3N^2}-{4 \over 3})+5) \cr
&\qquad \qquad \qquad \qquad \qquad \qquad \qquad \qquad \qquad
+ L_L (-{{(2e_L -N)^2} \over N^2}+4) \cr
&\qquad \qquad \qquad \qquad \qquad \qquad \qquad \qquad \qquad
+ L_E (-2{e_E^2 \over N^2})], \cr}
\equn\put\fsit
$$
where $L_R \equiv \sum_i \ln ({{M_{\rm GUT}} \over {m_{R_i}}})$ with the
sum running over all supermultiplets $i$ in the representation $R$.
$\alpha_s (m_{Z})$, $\alpha_{em}(m_{Z})$ and $\sin^2{\theta}_W(m_{Z})$ are
defined in the  $\overline{\rm MS}$ scheme, while the small numbers $-0.21$
and $0.0029$ in the r.h.s. of {\fmgut} and {\fsit} take into account two-loop
corrections [\caekn]. Substituting all $L_R = 0$, the above equations take
into account the particle spectrum of the minimal supersymmetric Standard
Model. In this case, one obtains good agreement with the experimental value
of the weak angle and a predicted value for ${M_{\rm GUT}}\sim 2\times
10^{16}{\rm GeV}$  [\clep]. This value turns out to be more than one order
of magnitude smaller than the string unification scale
$M_{\rm SU} \sim 3.73 \times 10^{17}{\rm GeV}$ [\cmsu].

The existence of additional representations, besides those of the minimal
supersymmetric Standard model, could move ${M_{\rm GUT}}$ close to
$M_{\rm SU}$ without destroying the succesful prediction for
$\sin^2{\theta}_W(m_{Z})$ if they satisfy the following inequalities:
$$
53 \simlt [{L_Q \over 6}(7-({{2e_Q -3N}\over N})^2 )
+ {L_D \over 3} (4-{e_D^2 \over N^2})
- {L_L \over 2} ({{(2e_L -N)^2} \over N^2}+1)
-  L_E ({e_E^2 \over N^2})] \simlt 70
\equn\put\nmgut
$$
\vskip 0.1truecm
$$
\eqalign{-22 \simlt [{L_Q \over 3} (-({{2e_Q -3N}\over N})^2 +22)
&- L_D (2({e_D^2 \over 3N^2}-{4 \over 3}) +5) \cr
&+ L_L (-{{(2e_L -N)^2} \over N^2}+4)
+ L_E (-2{e_E^2 \over N^2})] \simlt 7.5. }
\equn\put\nsit
$$
The above relations are obtained from {\fmgut} and {\fsit}, using the
values $M_{\rm GUT} = M_{\rm SU} \sim 3.73\times 10^{17}{\rm GeV}$,
$\alpha_{em}(m_{Z}) ={1 \over 128}$, $\sin^2{\theta}_W (m_{Z}) = 0.233$ and

$\alpha_s (m_{Z}) = 0.118 \pm 0.007$ [\cas]. The inequalities {\nmgut} and
{\nsit} were studied in Ref.[\caekn] for the case of non-exotic and
integrally charged Standard Model representations. These correspond to the
values  ${e_Q\over N}=2$, ${e_D\over N}=$ -2 or 1, ${e_L\over N}=1$ and
${e_E\over N}=-1$, which are associated to additional quark doublets, up
or down antiquarks, leptons and antileptons, respectively. In this case,
it was found that every solution of {\nmgut} and {\nsit} requires the
existence of at least one additional pair of quark doublets $Q+{\bar Q}$
with
mass much smaller  than $M_{\rm SU}$. This follows from the combination 3
$\times$ {\nmgut} + {\nsit}, which reads:
$$
27 \simlt [{L_Q\over 6}(13-({2e_Q\over N}-3)^2) +
{L_D\over 3}(1-{e_D^2\over N^2}) + {L_L\over 2}(1-({2e_L\over N}-1)^2)
- {e_E^2\over N^2}L_E ] \simlt 44.
\equn\put\ineq
$$
For the above mentioned values, the coefficients of all terms in {\ineq}
become negative except the one of $L_Q$, implying $L_Q\ne 0$. However, in
the presence of FEC-particles the coefficient of $L_L$ is also positive
and more general solutions exist [\cbl].

At the string unification scale $M_{\rm SU}$ the coupling constant of the
gauge group $G$ is also unified with the three Standard Model couplings and
is equal to $\alpha_{\rm GUT}$. This leads to an additional constraint
on the matter content of $G$, emerging from the requirement that the
confinement scale $\Lambda$ must have the desired value. A necessary
condition is that when taking into account just the FEC-states in the
renormalization group equation for the $G$ coupling constant in {\fbetag}
and {\fbi}, the resulting confinement scale $\Lambda_{\rm FEC}$ verifies:
$$
\Lambda_{\rm FEC} \simgt \delta M\sim{\rm TeV}.
\equn\put\lambda
$$
\vskip 1.0cm
\centerline{\bf NUMERICAL EXAMPLES}
\vskip 0.5cm

Here, we work out some explicit numerical examples to show that the
various masses of fractionally charged supermultiplets, chosen
appropriately, can lead to a reasonable spectrum of ordinary sparticles in
the region of a hundred GeV. At the same time we pay attention to be
consistent with the perturbative unification of all couplings at the string
scale, which implies in particular that the inequalities {\nmgut}, {\nsit}
and {\lambda} must hold.

The choice of the ``hidden" group $G$ has to satisfy the observed charge
quantization that singlet bound states of $SU(3)\times G$ must have integer
electric charges. For $\displaystyle{G=\prod_N SU(N)\times\prod_n SO(2n)}$,

this reduces to the following condition (for Kac-Moody
levels $k=1$) [\cSch],[\celn]:
$$
\sum_N{{i_N (N-i_N)}\over{2N}}+\sum_n \cases{0 &for $j_n =0$\cr
1/2 &for $j_n =2$\cr
n/8 &for $j_n =1$\cr}~={\rm non~ zero~ integer},
\equn\put\cqc
$$
where for every $N$, $i_N$ is some integer between 0 and $N-1$. Moreover,
the electric charge of a state transforming in the representation
$N$ or $\overline{N}$ of $SU(N)$ and/or $2n$ of $SO(2n)$ is given by:
$$
q= \pm\sum_N{i_N \over N}+\sum_n {j_n \over 2}~~{\rm mod}1,
\equn\put\cfc
$$
where the $\pm$ sign stands for the representations $N$ or $\overline{N}$,
and the integers $i_N$ and $j_n$ are those which satisfy {\cqc}.

In the case where $G$ is semi-simple, the smallest group satisfying {\cqc}
is
$SO(16)$ (with $j_8 =1$) for orthogonal groups or $SU(8)$ (with $i_8 =4$)
for the unitary ones. In both cases, the electric charges of FEC
color-singlet states are half-integers. For such big groups one must be
careful in the choice of the matter content, so that the value of
$\alpha_{\rm GUT}$ is in the perturbative domain. Notice that the presence
of even one representation of type $D+{\bar D}$, which is the minimum
requirement for non-zero one-loop gluino masses, leads to an $SU(3)$ beta
function $b_3 =N-3$ becoming large and positive for large N. Perturbative
unification then implies that the masses $M_D$ of these representations are
very high. For example, $M_D\simgt 10^{14}$ GeV for $SO(16)$, or $M_D\simgt
10^{10}$ GeV for $SU(8)$.

When $G$ is a product of semi-simple factors many more solutions exist.
Here, we examine an example, where $G=SU(4)\times SO(2n)$ (with $i_4 =2$
and $j_n =2$) which is a solution of {\cqc} for every $n$ with half-integer
electric charges for the FEC color-singlet states. For simplicity we will
assume that all FEC light states transform non-trivially only under
$SU(4)$,
which is for instance the case in the flipped $SU(5)$ model of Ref.[\cConf]
with hidden sector $SU(4)\times SO(10)$. In this case, perturbative
unification implies no bounds on the masses $M_D$.

Now, we derive the mass spectrum for an example with 
$SU(4)$ group. Using the values:
$$\eqalign{ 
&M_{SU} \sim 3.7 \times 10^{17} {\rm GeV} \quad
\delta M =15 {\rm TeV} \quad {\rm confinement}\quad {\rm group:}\, 
SU(4)\times G,\cr
&n_Q=1, e_Q=1, m_Q=3 \times 10^{14} {\rm GeV},\quad
n_D=2, e_D=1, m_D=2 \times 10^{15}{\rm GeV},\quad \cr
& n_L=0,\quad
n_E=1, e_E=3, m_E= 10^{13} {\rm GeV}. \cr}$$ we find:
\vskip 1.0truecm

\vbox {\tabskip=0pt \offinterlineskip\def\tablerule{\noalign{\hrule}}
\halign to350pt {\strut#& \vrule#\tabskip=1em plus2em& \hfil#& \vrule#&
 \hfil#& \vrule# \tabskip=0pt\cr\tablerule
&&\omit\hidewidth Sparticles \hidewidth&&
 \omit\hidewidth Masses (in GeV) \hidewidth&\cr\tablerule
&&gluinos&&298 &\cr\tablerule
&&charginos&&960 &\cr\tablerule
&& neutralinos&& 320 &\cr\tablerule
&&$\delta m_{\tilde u_L}$&&1350 &\cr\tablerule
&&$\delta m_{\tilde d_L}$&&1350 &\cr\tablerule
&&$\delta m_{\tilde u_R}$&&450 &\cr\tablerule
&&$\delta m_{\tilde d_R}$&&450 &\cr\tablerule
&&$\delta m_{\tilde e_L}$&&1290 &\cr\tablerule
&&$\delta m_{\tilde \nu_L}$&&1290 &\cr\tablerule
&&$\delta m_{\tilde e_R}$&&209 &\cr\tablerule\hfil\cr}}

As already mentioned above, the values of scalar masses do not include the
Higgs, as well as possible superpotential contributions, which are model
dependent. In the case of minimal supersymmetric standard model, the
Higgs contribution, neglecting Yukawa couplings, represents small
correction for heavy particles but potentially important correction for
the light ones. Yukawa couplings are expected to modify mainly the stop
masses.

A particular characteristic of this scenario is the large mass difference
between squarks and sleptons, which is a consequence of the radiative
origin
of supersymmetry breaking communication to the observable sector. This
difference is due to the ratio of the strong over the electroweak coupling
constant and could be compensated only by the addition of many
FEC-particles in representations of type $L$ and $E$. Another
characteristic of this mechanism is that the right-selectron seems to be
the lightest supersymmetric particle, although a more careful study of the
Higgs sector is required for a definite conclusion. Finally, this class of
models predicts new composite heavy particles with integer electric
charges. The explicit construction of 4D string models with the desired
properties described in this work remain of course an open problem.

\vskip 1.0cm
\centerline{\bf REFERENCES}
\vskip 0.5cm

\parskip=-3 pt

\item{[{\cWW}]} X. Wen and E. Witten, {\np} {\bf B261} (1985)
651.\hfill\break

\item{[{\cAth}]} G. Athanasiu, J. Atick, M. Dine and W. Fischler, {\pl}
{\bf
214B} (1988) 55.\hfill\break

\item{[{\cSch}]} A. Schellekens, {\pl} {\bf 237B} (1990) 363.\hfill\break

\item{[{\cAnt}]} I. Antoniadis, {\it Proceedings of Summer School in High
Energy Physics and Cosmology, Trieste} (1990) 677.\hfill\break

\item{[{\cCamb}]} D. Bailey, B. Campbell and S. Davidson, {\it Phys. Rev.}
{\bf D43} (1991) 2314.\hfill\break

\item{[{\cConf}]} I. Antoniadis, J. Ellis, J. Hagelin and D.V. Nanopoulos,
{\pl} {\bf 231B} (1989) 65.\hfill\break

\item{[{\cGC}]} J.P. Derendinger, L.E. Ib\'a\~nez and H.P. Nilles, {\pl}
{\bf 155B} (1985) 65;
M. Dine, R. Rohm, N. Seiberg and E. Witten, {\pl} {\bf 156B} (1985) 55.
\hfill\break

\item{[{\celn}]} J. Ellis, J.L. Lopez and D.V. Nanopoulos, {\pl} {\bf 245B}
(1990) 375 and {\bf 247B} (1990) 257.\hfill\break

\item{[{\cBarb}]} R. Barbieri, L. Girardello and A. Masiero, {\pl} {\bf
127B}
(1983) 429.\hfill\break

\item{[{\ck}]} K. Benakli, {\it Quelques aspects de la brisure de la 
supersym\'etrie en th\'eorie des cordes}, Ph.D Thesis, 1994;  M. Dine  
and A. E. Nelson, Phys. Rev. D48 (1993) 1277. \hfill\break

\item{[{\cil}]} L.E. Ib\'a\~nez and C. L\'opez, {\pl} {\bf 126B} (1983) 54;
{\np} {\bf B233} (1984) 511.\hfill\break

\item{[{\caekn}]} I. Antoniadis, J. Ellis, S. Kelley and D.V. Nanopoulos,
{\pl} {\bf 272B} (1991) 31.\hfill\break

\item{[{\clep}]} J. Ellis, S. Kelley and D.V. Nanopoulos, {\pl} {\bf 249B}
(1990) 441 and {\bf 260B} (1991) 131;
U. Amaldi, W.de Boer and H. F\"urstenau, {\pl} {\bf 260B} (1991) 447;
P. Langacker and M. Luo, {\it Phys. Rev.} {\bf D44} (1991) 817.\hfill\break

\item{[{\cmsu}]} V.S. Kaplunovsky, {\np} {\bf B307} (1988) 145 and {\it
Errata} STANFORD-ITP-838 preprint (1992).\hfill\break

\item{[{\cas}]} S. Bethke, {\bf XXVI} {\it International Conference on High
Energy Physics, Dallas, August} 1992.\hfill\break

\item{[{\cbl}]} D. Bailin and A. Love, {\it Mod. Phys. Lett.} {\bf A7}
(1992) 1485.\hfill\break

\end